# Nonlocal strain gradient torsion of elastic beams: variational formulation and constitutive boundary conditions

R. Barretta*, S. Ali Faghidian, F. Marotti de Sciarra, M. S. Vaccaro

*Department of Structures for Engineering and Architecture, University of Naples Federico II,*
*via Claudio 21, 80125 Naples, Italy*
e-mails: rabarret@unina.it - faghidian@gmail.com - marotti@unina.it - marziasara.vaccaro@unina.it

**Abstract**

Nonlocal strain gradient continuum mechanics is a methodology widely employed in literature to assess size effects in nanostructures. Notwithstanding this, improper higher-order boundary conditions (HOBC) are prescribed to close the corresponding elastostatic problems. In the present study, it is proven that HOBC have to be replaced with univocally determined boundary conditions of constitutive type, established by a consistent variational formulation. The treatment, developed in the framework of torsion of elastic beams, provides an effective approach to evaluate scale phenomena in smaller and smaller devices of engineering interest. Both elastostatic torsional responses and torsional free vibrations of nano-beams are investigated by applying a simple analytical method. It is also underlined that the nonlocal strain gradient model, if equipped with the inappropriate HOBC, can lead to torsional structural responses which unacceptably do not exhibit nonlocality. The presented variational strategy is instead able to characterize significantly peculiar softening and stiffening behaviours of structures involved in modern Nano-Electro-Mechanical-Systems (NEMS).

*Corresponding author
E-mail: rabarret@unina.it

# 1. Introduction

Nano-scale structures are extensively utilized in the design and manufacturing of modern nano-electro-mechanical systems (NEMS) due to their superior features [1]. Torsional nano-beams are fundamental structural elements of pioneering nano-engineered devices such as micro-thruster of nano-satellites [2], bio-torsional actuator [3], torsional accelerometer [4] and viscosity and mass density sensor [5]. It is well established that the local elasticity theory cannot be adopted to capture mechanical responses of nano-continua, and consequently, a variety of size-dependent elasticity models are available in literature. Analysis and assessment of size-effects in nano-structures is currently a topic of major interest in the scientific community [6-20]. Torsional deformations can frequently occur in structural elements of NEMS, and therefore, various size-dependent elasticity theories have been exploited in literature [21-32], as comprehensively discussed in review contributions [33, 34]. In fully nonlocal mechanics, elastic nonlocal fields are output of integral convolutions involving local source fields and suitable smoothing kernels. Two stress- and strain-driven nonlocal elasticity formulations are therefore accessible in literature. Eringen's strain-driven nonlocal law can be adopted to examine dislocations and wave propagation, which are problems formulated in unbounded domains [35], but not for nano-structural applications. Indeed, the constitutive boundary conditions associated with Eringen's integral convolution in nonlocal structural problems of engineering interest are in contrast with equilibrium requirements, and thus, the strain-driven law can lead to ill-posed elastic formulations [36]. The stress-driven nonlocal integral model, conceived for nano-beams in [37] and recently extended to axisymmetric nano-plates in [38], yields instead a mathematically well-posed and effective nonlocal approach in structural applications of nanotechnology. Pure and two-phase stress-driven nonlocal elasticities have been applied in a series of papers to study elastostatic responses [28, 29, 39-43], free vibrations [44-48] and stability of nano-beams [49].

To account for size-dependent behavior of nano-continua, alternatively, mechanical responses can be assumed to depend on both strain fields and gradient fields of higher-order. The strain gradient elasticity incorporates all components of the higher-order deformation, and thus, it is too intricate to be exploited in the original framework. A variety of simplified versions of strain gradient constitutive models have been introduced in literature and employed to address a range of structural problems. Instances are the following: most general strain gradient theory [25, 27, 50], generalized first strain gradient elasticity theory [26, 51], second strain gradient elasticity theory [52], strain gradient elasticity with surface energy [53, 54], modified strain gradient elasticity [55-59] and modified couple stress theory [60-62]. Notably, different forms of strain gradient models lead however to stiffening structural responses for increasing small-scale parameters [63]. Interestingly, Eringen's nonlocal differential law and the strain gradient elasticity model were coupled by Aifantis in [64, 65], proposing thus a nonlocal strain gradient elasticity theory. Eringen's nonlocal integral relation was then combined with the strain gradient elasticity by Lim et al. [66] to describe a wider variety of higher-order nonlocal strain gradient materials. In this framework, the governing differential constitutive law is of higher-order than the classical local one due to gradient effects. To close the nonlocal strain gradient problem, Lim et al. [66] introduced higher-order boundary conditions (HOBC), overlooking the constitutive boundary conditions (CBC) associated with the integral convolutions, recently established by Barretta & Marotti de Sciarra [67] for Bernoulli-Euler inflected nano-beams nd exploited in [68-70]. Hereafter, the nonlocal strain gradient model with CBC will be labeled by MNSG, standing for modified nonlocal strain gradient. It is worth underlining that, nonlocal problems of wave propagation (defined in unbounded domains) were studied by Lim et al. in [66]. In that context, HOBC and CBC were appropriately not prescribed being tacitly verified, due to the fact that the fields involved in the integral convolutions are rapidly vanish at infinity.

Structural problems of technical interest are instead defined in bounded domains and, therefore, appropriate CBC have to be imposed to close the constitutive formulation.

## 2. Motivation and outline

In a recent paper [71], it is claimed that both HOBC and CBC have to be prescribed in structural mechanics, with the unpleasant conclusion that the elastostatic problem of a nonlocal strain gradient beam under flexure generally admits no solution in nano-engineering. In the framework of torsion of nano-beams, a main motivation of the present paper is to prove that, if the nonlocal strain gradient model of elasticity is properly formulated in variational terms, the relevant elastostatic and elastodynamic problems are well-posed. Secondly, it is shown that HOBC, usually applied by the scientific community [31, 32], have nothing to do with the nonlocal strain gradient model of elasticity and that they have to be replaced with the torsional CBC, here established for the first time, which are variationally consistent. While inherent ill-posedness of the strain-driven nonlocal elasticity is well-discussed in literature, the strain gradient theory can lead to well-posed problems. However, the strain gradient continuum theory is well-recognized to exhibit merely stiffening structural behaviors at nano-scales. Contrary to the peculiar smaller-is-stiffer trend associated with strain gradient-type theories, the conceived modified nonlocal strain gradient theory is expediently able to predict both stiffening and softening structural responses at small-scales. Due to the significant influence of the prescribed non-classical boundary conditions on the torsional response of nano-beams, in the present study it is firstly re-examined the elastic torsion in the framework of modified nonlocal strain gradient model employing the appropriate constitutive variational formulation, which leads to the consistent form of the non-classical CBCs to be imposed to close the relevant structural problem. In Sect. 3, torsional behaviour of nonlocal strain gradient elastic nano-beams of technical interest, involving bounded domains and standard kinematic boundary constraints, are variationally evaluated by constitutive integral convolutions which will be properly replaced with a differential problem, equipped with the constitutive boundary conditions to be prescribed. Static and dynamic torsional elastic responses of nano-beams are then investigated in Sect. 4 and 5, respectively. The contributed results are also compared with pertinent outcomes of the known nonlocal strain gradient model, equipped with higher-order boundary conditions. Drawbacks and difficulties of the classical nonlocal strain gradient theory, in consequence of prescribing the unmotivated HOBC, are elucidated and discussed. New numerical illustrations are presented for the elastostatic torsional response and fundamental torsional frequency of nano-beams that can be exploited for structural design of NEMS. Closing remarks are outlined in Sect.6.

## 3. Modified nonlocal strain gradient law for torsion

A straight homogeneous isotropic nano-beam of length $L$, material density $\rho$ with circular cross-sectional domain $\Xi$ is considered and schematically depicted in Fig. 1. The abscissa $x$ is taken along the length of the nano-beam which is orthogonal to the plane of the cross-section containing the position vector $\mathbf{r}$ with respect to the cross-section centroid. Elasticity solution of Saint-Venant's problem [72] is utilized to set forth the formulation of the beam under torsion. The displacement field for the torsion of nano-beam $\mathbf{u}$, up to an inessential additional rigid-body motion, writes thus as

$$\mathbf{u}(\mathbf{r},x) = \theta(x)\mathbf{R}\mathbf{r} \qquad (1)$$

where the tensor $\mathbf{R}$ is the rotation by $\pi/2$ counterclockwise in $\Xi$ and $\theta$ stands for the torsional rotation field. The ensuing kinematically compatible shear strain is provided by

$$\boldsymbol{\gamma}(\mathbf{r},x) = \chi_t(x)\mathbf{R}\mathbf{r} \qquad (2)$$

where $\boldsymbol{\gamma}$ is the shear strain vector and $\chi_t = \partial_x \theta$ stands for geometric torsional curvature. Accordingly, the local shear stress field $\boldsymbol{\tau}$ associated with the shear strain $\boldsymbol{\gamma}$ takes the form

$$\boldsymbol{\tau}(\mathbf{r},x) = \mu\boldsymbol{\gamma}(\mathbf{r},x) = \mu(\partial_x \theta)\mathbf{R}\mathbf{r} \tag{3}$$

with $\mu$ shear elastic modulus. The loading condition of the nano-beam consists of a distributed couple function $m_t$ in $[0,L]$ and concentrated couples $\bar{M}_t$ at the beam ends. Differential and classical boundary conditions of dynamic equilibrium write as

$$\begin{aligned}\partial_x M_t + m_t &= J_\rho \partial_{tt}\theta \\ (M_t + \bar{M}_t)\delta\theta\big|_{x=0} &= (M_t - \bar{M}_t)\delta\theta\big|_{x=L} = 0\end{aligned} \tag{4}$$

where $J_\rho$ and $M_t$, respectively, are the mass polar moment of inertia and the twisting moment expressed by

$$J_\rho = \iint_\Xi \rho(\mathbf{r}\cdot\mathbf{r})dA, \qquad M_t = \iint_\Xi (\mathbf{R}\mathbf{r})\cdot\boldsymbol{\tau}\,dA \tag{5}$$

with the dot into the integral denoting the inner product between vectors.

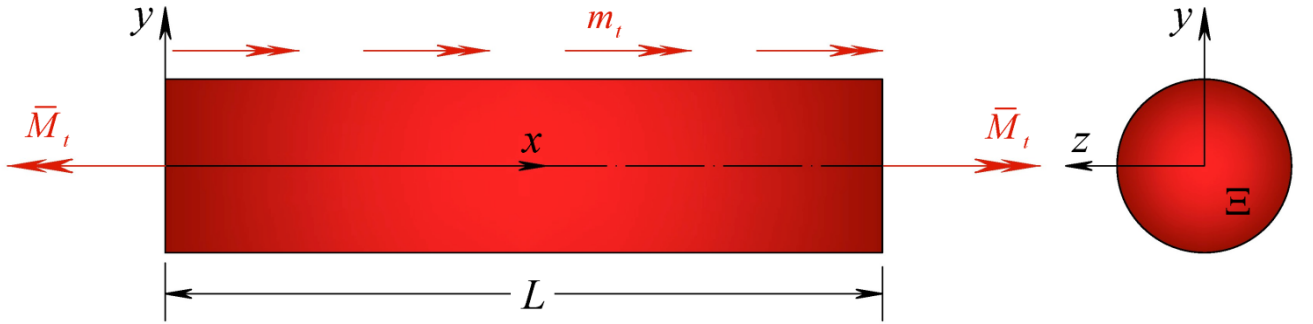

**Fig. 1**. Coordinate system and configuration of a nano-beam under torsion.

In strain gradient theory (SGT), the elastic strain energy dependent on the elastic torsional curvature $\chi_t \in C^2([0,L];\Re)$ writes as [28]

$$\Pi_{SGT}(\chi_t) := \frac{1}{2}\int_0^L \left(J_\mu \chi_t^2 + J_\mu \ell^2 (\partial_x \chi_t)^2\right)dx \tag{6}$$

where the gradient characteristic length $\ell$ is introduced to establish the significance of the first-order strain gradient field. The local elastic torsional stiffness $J_\mu$ is the polar moment of the field of shear elastic modulus $\mu$ as

$$J_\mu = \iint_\Xi \mu(\mathbf{r}\cdot\mathbf{r})dA \tag{7}$$

which is assumed to be independent of the $x$ coordinate. Stimulated by the seminal approach of Eringen [35], the torsional problem is formulated in the framework of modified nonlocal strain gradient theory (MNSG). As a result of suitably introducing nonlocal integral convolutions in Eq. (6), the elastic strain energy $\Pi_{MNSG}$ of a straight beam under torsion, associated with MNSG, is proposed as

$$\Pi_{MNSG}(\chi_t) := \frac{1}{2}\int_0^L \left(J_\mu (\varphi_0 * \chi_t)\chi_t + J_\mu \ell^2 (\varphi_1 * \partial_x \chi_t)(\partial_x \chi_t)\right)dx \tag{8}$$

where the smoothing kernels $\varphi_0$ and $\varphi_1$ depend on two non-dimensional nonlocal parameters $\lambda_0 > 0$ and $\lambda_1 > 0$. The definition of integral convolution of a scalar source field $s$ with a scalar kernel $\varphi_i$ is also recalled as

$$(\varphi_i * s)(x,\lambda_i) := \int_0^L \varphi_i(x - \bar{x}, \lambda_i)s(\bar{x})d\bar{x} \tag{9}$$

with $x$ and $\bar{x}$ representing points of $[0,L]$. The scalar kernels $\varphi_0, \varphi_1$ are assumed to fulfill positivity, parity, symmetry, normalization and limit impulsivity properties [36].

The stress field in an elastic nano-beam under torsion associated with MNSG is the twisting moment $M_t \in C^1([0,L];\Re)$ that can be determined by the variational constitutive condition

$$\langle M_t, \delta\chi_t \rangle := \int_0^L M_t(x)\delta\chi_t(x)dx = \langle d\Pi_{MNSG}(\chi_t), \delta\chi_t \rangle \tag{10}$$

for any virtual torsional curvature field $\delta\chi_t \in C_0^1([0,L];\Re)$ having compact support in $[0,L]$. The directional derivative of the elastic strain energy along a virtual torsional curvature is then established introducing the expression of $\Pi_{MNSG}$ Eq. (8), while integrating by parts

$$\begin{aligned}\langle d\Pi_{MNSG}(\chi_t), \delta\chi_t \rangle &= \int_0^L \left( J_\mu(\varphi_0 * \chi_t)\delta\chi_t + J_\mu \ell^2 (\varphi_1 * \partial_x \chi_t)(\partial_x \delta\chi_t) \right)dx \\ &= \int_0^L \left( J_\mu(\varphi_0 * \chi_t) - J_\mu \ell^2 \partial_x(\varphi_1 * \partial_x \chi_t) \right)\delta\chi_t dx \\ &\quad + J_\mu \ell^2 \left( (\varphi_1 * \partial_x \chi_t)\delta\chi_t \big|_{x=L} - (\varphi_1 * \partial_x \chi_t)\delta\chi_t \big|_{x=0} \right)\end{aligned} \tag{11}$$

The test fields $\delta\chi_t \in C_0^1([0,L];\Re)$ in the variational constitutive condition Eq. (10) are assumed to have compact support, so that the boundary values $\delta\chi_t|_{x=0}$ and $\delta\chi_t|_{x=L}$ are vanishing, and accordingly, the boundary terms in Eq. (11) disappear. The twisting moment $M_t$ is then determined in terms of the torsional curvature $\chi_t$ as a result of applying a standard localization procedure while implementing the variational condition Eq. (10)

$$\begin{aligned}M_t &= J_\mu(\varphi_0 * \chi_t)(x,\lambda_0) - J_\mu \ell^2 \partial_x \left( (\varphi_1 * \partial_x \chi_t)(x,\lambda_1) \right) \\ &= J_\mu \int_0^L \varphi_0(x-\bar{x},\lambda_0)\chi_t(\bar{x},t)d\bar{x} - J_\mu \ell^2 \partial_x \int_0^L \varphi_1(x-\bar{x},\lambda_1)(\partial_{\bar{x}}\chi_t)(\bar{x},t)d\bar{x}\end{aligned} \tag{12}$$

Due to the intricacy of the formulations from physical and mathematical point of view, it is normally assumed that the nonlocal parameters are coincident, $\lambda_0 = \lambda_1 = \lambda > 0$ [31, 32, 66]. Furthermore, the Helmholtz bi-exponential smoothing kernel is a common choice in literature for the smoothing kernels $\varphi_0$ and $\varphi_1$ given by

$$\varphi(x,L_c) = \frac{1}{2L_c}\exp\left(-\frac{|x|}{L_c}\right) \tag{13}$$

with $L_c = \lambda L$ representing the characteristic length of Eringen nonlocal elasticity demonstrating long-range interactions. Helmholtz bi-exponential kernel is well-established to fulfill positivity, symmetry, normalization and impulsivity conditions [36].

Accordingly, the twisting moments $M_t$ can be rewritten as

$$\begin{aligned} M_t &= J_\mu(\varphi * \chi_t)(x,\lambda) - J_\mu \ell^2 \partial_x\left((\varphi * \partial_x \chi_t)(x,\lambda)\right) \\ &= J_\mu \int_0^L \varphi(x-\bar{x},\lambda)\chi_t(\bar{x},t)d\bar{x} - J_\mu \ell^2 \partial_x \int_0^L \varphi(x-\bar{x},\lambda)(\partial_{\bar{x}}\chi_t)(\bar{x},t)d\bar{x} \end{aligned} \tag{14}$$

Following the approach in Prop. 3.1 by Barretta and Marotti de Sciarra [67], the nonlocal strain gradient convolutions Eq. (14) can be shown to be equivalent to an appropriate differential constitutive law equipped with suitable constitutive boundary conditions.

**Proposition 1. Constitutive equivalency for torsional nano-beams**

The modified nonlocal strain gradient constitutive law Eq. (14) with the bi-exponential kernel Eq. (13) for torsional nano-beams defined on a bounded interval $[0,L]$ is equivalent to the differential constitutive relation

$$M_t(x,t) - L_c^2 \partial_{xx} M_t(x,t) = J_\mu \chi_t(x,t) - J_\mu \ell^2 \partial_{xx}\chi_t(x,t) \tag{15}$$

subjected to constitutive boundary conditions (CBCs) at the beam ends $x = 0$ and $x = L$ as

$$\begin{aligned} \partial_x M_t(0,t) - \frac{1}{L_c}M_t(0,t) &= \frac{\ell^2}{L_c^2}J_\mu \partial_x \chi_t(0,t) \\ \partial_x M_t(L,t) + \frac{1}{L_c}M_t(L,t) &= \frac{\ell^2}{L_c^2}J_\mu \partial_x \chi_t(L,t) \end{aligned} \tag{16}$$

**Remark 1.** By virtue Proposition 1 and due to the presence of the gradient term at r.h.s of Eq.(16), which is characterized by the non-vanishing parameter $\ell$, the differential equation Eq.(15) with CBC Eq. (16) associates univocally the torsional nonlocal strain gradient elastic curvature $\chi_t$ (source of the integral law Eq. (14)) with any fixed equilibrated twisting moment field $M_t$, known term of the integral law Eq. (14) and solution of the static problem Eq.(4). Nonlocal strain gradient torsional rotations $\theta$ are finally obtained by integrating the differential condition of kinematic compatibility $\chi_t = \partial_x \theta$ and enforcing standard (essential) kinematic boundary conditions. Notably, if the gradient parameter $\ell$ is assumed to vanish, the nonlocal strain gradient integral law Eq. (14) collapses into the purely nonlocal strain driven integral model by Eringen. CBCs Eq. (16) become therefore homogeneous conditions which are in contrast with standard (natural) static boundary conditions of structural problems of technical interest. A detailed discussion on ill-posedness of Eringen nonlocal theory in mechanics of beams under torsion is provided in [28]. Presence of the non-vanishing gradient parameter $\ell$ in Eq. (14) guarantees that analytical shapes of stress fields generated by the integral relation Eq. (14) are compatible with those dictated by equilibrium requirements. Evidences of this conclusion are elucidated in Sect. 4 and 5 by also examining static and dynamic nonlocal gradient responses of simple structural schemes in nano-engineering. Unlike Eringen's strain-driven nonlocal integral elasticity formulations, the nonlocal strain gradient model governed by convolutions Eq. (14) leads to well-posed structural problems.

**Remark 2.** As expected, the strain gradient formulation of elastic torsion can be recovered as the nonlocal characteristic length approaches zero. Recalling the limit impulsivity property of smoothing kernel $\varphi$, Eq. (14) accordingly gives the strain gradient differential equation

$$M_t(x,t) = J_\mu \chi_t(x,t) - J_\mu \ell^2 \partial_{xx} \chi_t(x,t) \tag{17}$$

with CBC detected via simplifying Eq. (16)

$$J_\mu \ell^2 \partial_x \chi_t (0,t) = 0$$
$$J_\mu \ell^2 \partial_x \chi_t (L,t) = 0 \tag{18}$$

It should be noted that to establish the modified nonlocal strain gradient model, nonlocal effects are consistently associated with the strain gradient elasticity model by a proper constitutive variational formulation. Zero- and first-order twisting moments are not employed in detecting the constitutive law involving twisting moments $M_t$ consistent with MNSG. Therefore, the unmotivated higher-order boundary conditions (HOBC) of strain gradient theory are not required to be prescribed. In nonlocal strain gradient theory (NSG) by Lim et al. [66], zero- and first-order twisting moments, $M_t^{(0)}$ and $M_t^{(1)}$, are instead introduced as

$$M_t^{(0)} = \int_0^L J_\mu \varphi_0 (x - \overline{x}, \lambda_0) \chi_t (\overline{x},t) d\overline{x}$$
$$M_t^{(1)} = \ell^2 \int_0^L J_\mu \varphi_1 (x - \overline{x}, \lambda_1) \partial_{\overline{x}} \chi_t (\overline{x},t) d\overline{x} \tag{19}$$

and the twisting moments $M_t$ given by

$$M_t = M_t^{(0)} - \partial_x M_t^{(1)} \tag{20}$$

For bounded domains, replacing improperly the convolutions Eq. (20) with a differential law [66], the principle of minimum total potential energy provides higher-order boundary conditions of strain gradient theory as introduced in Eq. (18) [31-32]

$$\partial_x \chi_t \big|_{x=0,L} = \partial_{xx} \theta \big|_{x=0,L} = 0 \tag{21}$$

implicating vanishing of the derivative of torsional curvature at the end cross-sections. Prescription of inapt higher-order boundary conditions Eq. (21) leads to an over-constrained [71] and inconsistent boundary value problem [67] in nonlocal strain gradient theory. These inconsistencies are eliminated by adopting the suitable variational formulation Eq. (10) exploited by analyzing both static responses and free vibrations of nano-beams under torsion.

## 4. Elastostatic torsional analysis

The elastostatic torsional response of fully-clamped and cantilever nano-beams of length $L$ subjected to a uniform distributed couple function $\bar{m}_t$ per unit length is examined here adopting modified nonlocal strain gradient and nonlocal strain gradient theories. The non-dimensional parameters: axial abscissa $\bar{x}$, nonlocal characteristic parameter $\lambda_c$, gradient characteristic parameter $\lambda_\ell$ and torsional rotation $\bar{\theta}$ are introduced as

$$\bar{x} = \frac{x}{L}, \qquad \lambda_c = \frac{L_c}{L}, \qquad \lambda_\ell = \frac{\ell}{L}, \qquad \bar{\theta}(\bar{x}) = \theta(x)\frac{J_\mu}{\bar{m}_t L^2} \tag{22}$$

Due to absence of the inertia terms in the elastostatic analysis, the differential condition of equilibrium Eq. (4)$_1$ can be integrated to detect the twisting moment $M_t$ in terms of an integration constant $\Lambda_1$ as

$$M_t(x) = -\int_0^x m_t(\zeta)d\zeta + \Lambda_1 \tag{23}$$

The constitutive differential equation Eq. (15) can be subsequently solved to determine the torsional curvature field $\chi_t$ in terms of integration constants $\Lambda_2, \Lambda_3$

$$\chi_t(x) = \Lambda_2 \exp\left(-\frac{x}{\ell}\right) + \Lambda_3 \exp\left(\frac{x}{\ell}\right) + \frac{1}{2\ell J_\mu}\exp\left(-\frac{x}{\ell}\right)\int_0^x \exp\left(\frac{\xi}{\ell}\right)\left(M_t(\xi) - L_c^2 \partial_{\xi\xi} M_t(\xi)\right)d\xi$$
$$- \frac{1}{2\ell J_\mu}\exp\left(\frac{x}{\ell}\right)\int_0^x \exp\left(-\frac{\eta}{\ell}\right)\left(M_t(\eta) - L_c^2 \partial_{\eta\eta} M_t(\eta)\right)d\eta \tag{24}$$

The torsional rotation field $\theta$ can be evaluated integrating the differential condition of kinematic compatibility $\chi_t = \partial_x \theta$ in terms of the integration constant $\Lambda_4$ as

$$\theta(x) = \int_0^x \chi_t(\zeta)d\zeta + \Lambda_4 \tag{25}$$

The integration constants $\Lambda_k (k=1..4)$ can be determined by prescribing the standard kinematic and static boundary conditions (BC) as well as the CBC. The proposed analytical approach provides exact solutions as a result of integrating differential equations of lower order. In the sequel, the acronyms LOC, NSG and MNSG, respectively, denote the local beam model, nonlocal strain gradient model and modified nonlocal strain gradient theory.

4.1. *Fully-clamped nano-beam subject to uniformly distributed couples*

The kinematic BCs in case of a fully-clamped nano-beam are given by

$$\theta(0) = 0, \qquad \theta(L) = 0 \qquad (26)$$

The non-dimensional torsional rotation field can be determined exploiting the proposed solution methodology while imposing the kinematic BCs Eq. (26) and the corresponding CBCs of MNSG Eq. (16) and NSG Eq. (21) as

$$\bar{\theta}^{\text{MNSG}}(\bar{x}) = \bar{\theta}^{\text{LOC}}(\bar{x}) - \frac{1}{2}\left(\lambda_c + 2\lambda_c^2 - 2\lambda_\ell^2\right)\left(-1 + \cosh\left(\frac{1-2\bar{x}}{2\lambda_\ell}\right)\text{sech}\left(\frac{1}{2\lambda_\ell}\right)\right)$$

$$\bar{\theta}^{\text{NSG}}(\bar{x}) = \bar{\theta}^{\text{LOC}}(\bar{x}) + \lambda_\ell^2\left(-1 + \cosh\left(\frac{1-2\bar{x}}{2\lambda_\ell}\right)\text{sech}\left(\frac{1}{2\lambda_\ell}\right)\right) \qquad (27)$$

where the local non-dimensional rotation field of fully-clamped nano-beam is well-established to write as

$$\bar{\theta}^{\text{LOC}}(\bar{x}) = \frac{1}{2}\bar{x}(1-\bar{x}) \qquad (28)$$

It can be inferred from Eq. (27) that the elastic torsional response of fully-clamped nano-beam subjected to uniformly distributed couples in the framework of NSG is independent of the nonlocal characteristic parameter $\lambda_c$.

The maximum torsional rotation field is also detected for numerical illustrations

$$\bar{\theta}_{\max}^{\mathrm{MNSG}} = \frac{1}{8}\left(1 + 4\lambda_c + 8\lambda_c^2 - 8\lambda_\ell^2 - 4\left(\lambda_c + 2\lambda_c^2 - 2\lambda_\ell^2\right)\mathrm{sech}\left(\frac{1}{2\lambda_\ell}\right)\right)$$

$$\bar{\theta}_{\max}^{\mathrm{NSG}} = \frac{1}{8} - \lambda_\ell^2 + \lambda_\ell^2 \mathrm{sech}\left(\frac{1}{2\lambda_\ell}\right) \tag{29}$$

### 4.2. *Cantilever nano-beam subject to uniformly distributed couples*

For a cantilever nano-beam subjected to uniformly distributed couples, the classical BCs are well-established to be

$$\theta(0) = 0, \qquad M_t(L) = 0 \tag{30}$$

Employing the solution technique introduced above, while prescribing the classical BCs in addition to the appropriate CBCs, the non-dimensional torsional rotation field of the nano-beam can be shown to have the expressions of

$$\bar{\theta}^{\mathrm{MNSG}}(\bar{x}) = \bar{\theta}^{\mathrm{LOC}}(\bar{x}) - 2\exp\left(\frac{1}{\lambda_\ell}\right)\left(-1 + \coth\left(\frac{1}{\lambda_\ell}\right)\right)\sinh\left(\frac{\bar{x}}{2\lambda_\ell}\right)\left(\left(\lambda_c^2 - \lambda_\ell^2\right)\cosh\left(\frac{\bar{x}}{2\lambda_\ell}\right)\right.$$
$$\left. - \left(\lambda_c + \lambda_c^2 - \lambda_\ell^2\right)\cosh\left(\frac{-2+\bar{x}}{2\lambda_\ell}\right)\right) \tag{31}$$

$$\bar{\theta}^{\mathrm{NSG}}(\bar{x}) = \bar{\theta}^{\mathrm{LOC}}(\bar{x}) + \lambda_\ell^2\left(-1 + \cosh\left(\frac{1-2\bar{x}}{2\lambda_\ell}\right)\mathrm{sech}\left(\frac{1}{2\lambda_\ell}\right)\right)$$

with the local non-dimensional rotation field of cantilever nano-beam expressed as

$$\bar{\theta}^{\mathrm{LOC}}(\bar{x}) = \frac{1}{2}(2 - \bar{x})\bar{x} \tag{32}$$

Once more, it can be deduced from Eq. (31) that NSG cannot capture the nonlocality effects in case of a cantilever nano-beam subjected to uniformly distributed couples.

For numerical illustration sake, the maximum torsional rotation field at the free end is evaluated as

$$\begin{aligned}\bar{\theta}_{max}^{MNSG} &= \frac{1}{2} + \lambda_c \\ \bar{\theta}_{max}^{NSG} &= \frac{1}{2}\end{aligned} \qquad (33)$$

While the torsional rotation of a MNSG nano-beam at the free end is independent of the gradient characteristic parameter $\lambda_\ell$, the maximum torsional rotation of NSG nano-beam coincides with the results of local elastic model of beams. Therefore to investigate the effects of nonlocal and gradient characteristic parameters, the value of the torsional rotation field at the mid-span of the beam is examined

$$\begin{aligned}\bar{\theta}^{MNSG}\left(\bar{x}=\frac{1}{2}\right) &= \frac{1}{8}\left(3+8\lambda_c(1+\lambda_c)-8\lambda_\ell^2-4\left(\lambda_c+2\lambda_c^2-2\lambda_\ell^2\right)\operatorname{sech}\left(\frac{1}{2\lambda_\ell}\right)\right) \\ \bar{\theta}^{NSG}\left(\bar{x}=\frac{1}{2}\right) &= \frac{3}{8}-\lambda_\ell^2+\lambda_\ell^2\operatorname{sech}\left(\frac{1}{2\lambda_\ell}\right)\end{aligned} \qquad (34)$$

*4.3. Numerical illustrations of elastostatic torsional response*

The normalized torsional rotation at the mid-span of the fully-clamped and cantilever nano-beams associated with the modified nonlocal strain gradient theory under uniformly distributed couples is exhibited in Figs. 2-3 in comparison with the counterpart results of the nonlocal strain gradient model. Detected torsional rotation fields are also normalized employing the corresponding torsional rotation of the local beam model $\bar{\theta}^{LOC}$. In Figs. 2-3 while the nonlocal characteristic parameter $\lambda_c$ is ranging in the interval $]0, 0.5]$, the gradient characteristic parameter $\lambda_\ell$ is ranging in the set of $\{0.1, 0.2, 0.3, 0.4, 0.5\}$. Furthermore, 3D plots of variations of normalized torsional rotation at the mid-span of the nano-beam versus the nonlocal and gradient characteristic parameters, $\lambda_c$ and $\lambda_\ell$, are illustrated in Figs. 4 and 5 for uniformly loaded fully-clamped and cantilever nano-beams. In Figs. 4-5 both the nonlocal and gradient characteristic parameters are ranging in the interval $]0, 0.5[$.

It is inferred from Figs. 2-5 that the torsional rotation field of nano-beam decreases as the gradient characteristic parameter $\lambda_\ell$ increases. Therefore, both the adopted nonlocal strain gradient models exhibit a stiffening behavior in terms of the gradient parameter $\lambda_\ell$ for a given value of $\lambda_c$. The nonlocal strain gradient model noticeably cannot capture nonlocality effects for both uniformly loaded fully-clamped and cantilever nano-beams. On the contrary, the modified nonlocal strain gradient theory efficiently demonstrates a softening behavior in terms of the nonlocal characteristic parameter $\lambda_c$ that is a larger $\lambda_c$ involves a larger torsional rotation for a given value of $\lambda_\ell$. While the demonstrated results in Figs. 2-5 reveal that the nonlocal strain gradient model is unable of appropriately assessing nonlocal size-effects in uniformly loaded torsional nano-beams, such a peculiar behavior is absent in the torsional results consistent with the modified nonlocal strain gradient model. As the nonlocal and gradient parameters approach zero $\lambda_c, \lambda_\ell \to 0^+$, the elastostatic torsional response of nano-beam associated with either of nonlocal strain gradient models is coincident with the local torsional rotation. The numerical values of normalized torsional rotation at the beam mid-span detected by NSG and MNSG for uniformly loaded fully-clamped and cantilever nano-beams are collected in Tables 1 and 2, respectively.

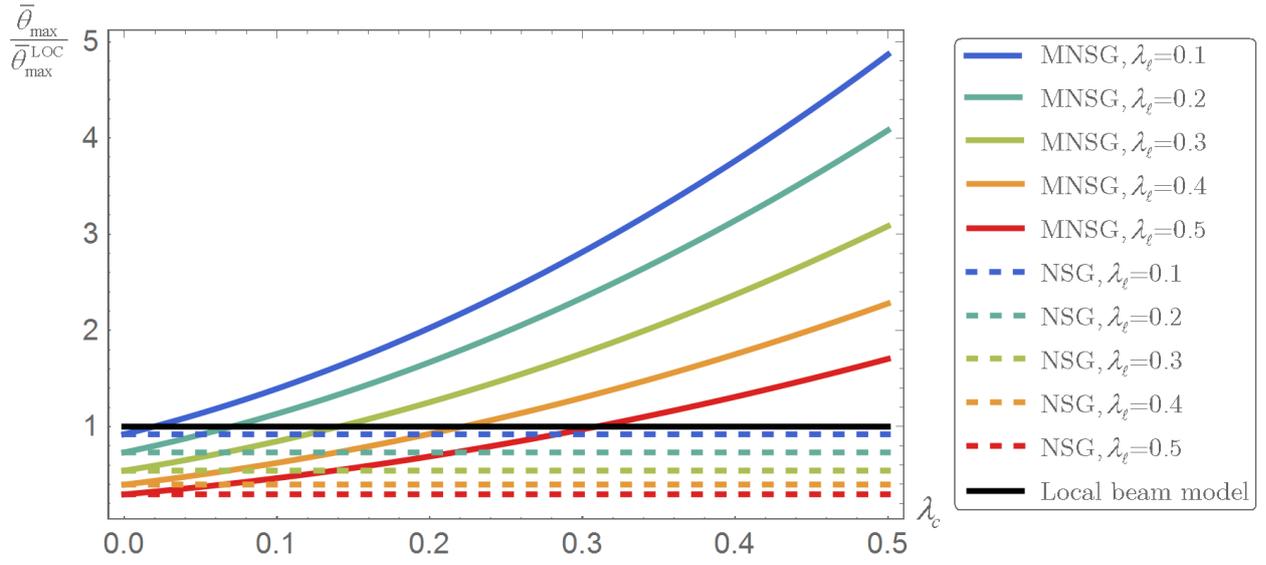

**Fig. 2.** Uniformly loaded fully-clamped nano-beam: effects of $\lambda_c, \lambda_\ell$ on normalized $\bar{\theta}_{max}$

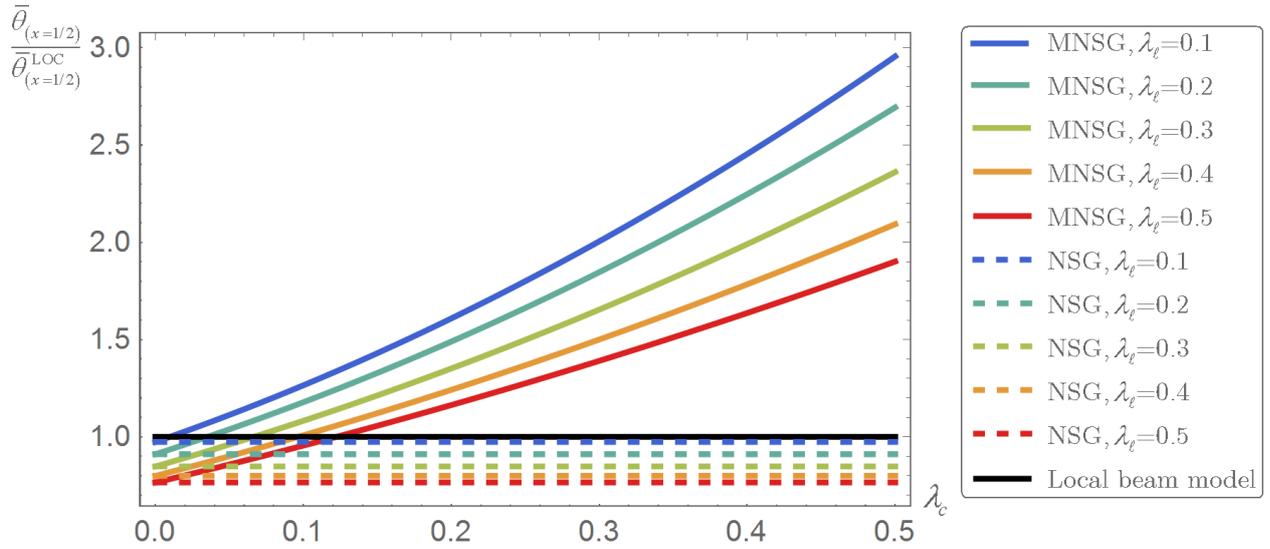

**Fig. 3.** Uniformly loaded cantilever nano-beam: effects of $\lambda_c, \lambda_\ell$ on normalized $\bar{\theta}_{(\bar{x}=1/2)}$

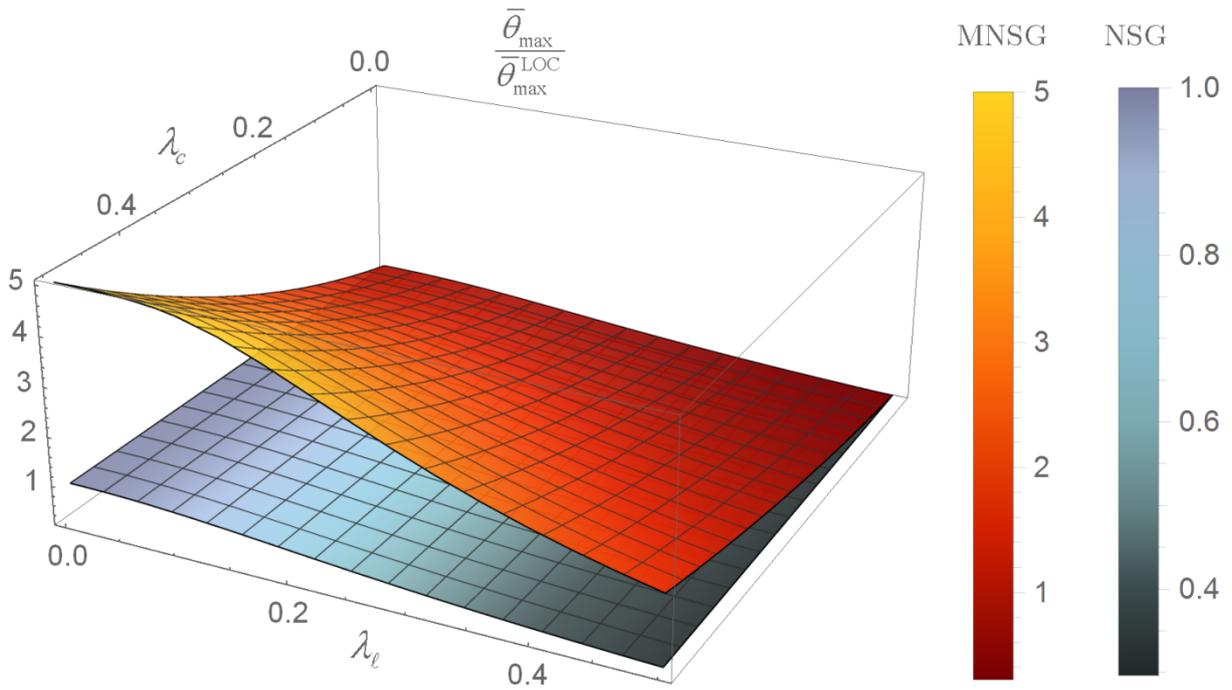

**Fig. 4.** Fully-clamped nano-beam under uniform couples: normalized $\bar{\theta}_{max}$ vs. $\lambda_c$ and $\lambda_\ell$

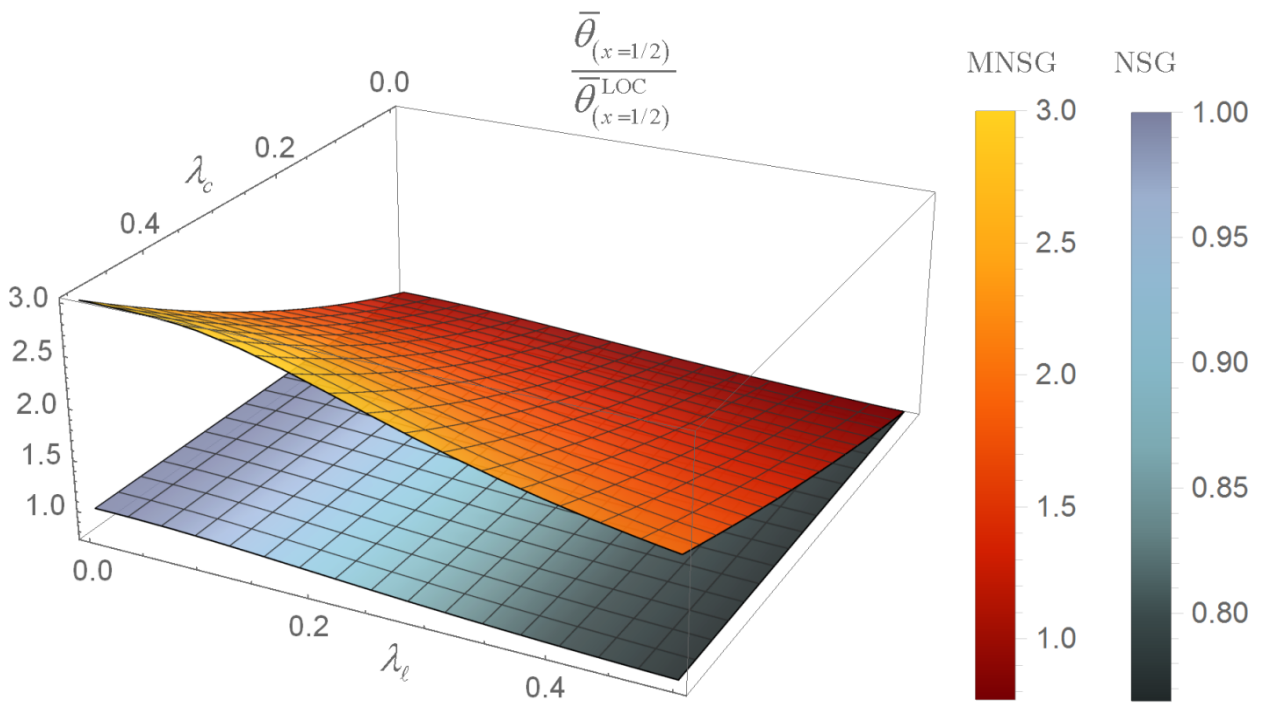

**Fig. 5.** Cantilever nano-beam under uniform couples: normalized $\bar{\theta}_{(\bar{x}=1/2)}$ vs. $\lambda_c$ and $\lambda_\ell$

## 5. Torsional free vibration analysis

In order to study torsional free vibrations of nano-beams, distributed couples are vanishing. Twisting moment fields $M_t$ can be therefore evaluated by imposing the differential condition of equilibrium Eq. (4)$_1$ to the constitutive differential law Eq. (15) as

$$M_t(x,t) = L_c^2 \partial_x (J_\rho \partial_{tt} \theta)(x,t) + J_\mu \chi_t(x,t) - J_\mu \ell^2 \partial_{xx} \chi_t(x,t) \tag{35}$$

The differential condition of dynamic equilibrium governing torsional vibrations of nano-beams can be expressed in terms of torsional rotation field by applying the kinematic compatibility $\chi_t = \partial_x \theta$

$$\partial_x (J_\mu \partial_x \theta)(x,t) - \ell^2 \partial_x (J_\mu \partial_{xxx} \theta)(x,t) = J_\rho \partial_{tt} \theta(x,t) - L_c^2 \partial_{xx} (J_\rho \partial_{tt} \theta)(x,t) \tag{36}$$

equipped with the classical boundary conditions Eq. (4)$_2$ and corresponding constitutive boundary conditions associated with MNSG Eq. (16) or NSG Eq. (21). A standard procedure of separating spatial and time variables is then utilized to study torsional free vibrations

$$\theta(x,t) = \Theta(x)\exp(i\omega t) \tag{37}$$

with $i = \sqrt{-1}$, $\Theta$ and $\omega$ representing the spatial mode shapes and natural frequency of torsional vibrations. Imposing the separation of variables Eq. (37) on the differential conditions of dynamic equilibrium Eq. (36), the differential condition of torsional coordinate functions $\Theta$ writes as

$$\ell^2 d_x (J_\mu d_{xxx} \Theta)(x) - d_x (J_\mu d_x \Theta)(x) + \omega^2 L_c^2 d_{xx}(J_\rho \Theta)(x) - \omega^2 J_\rho \Theta(x) = 0 \tag{38}$$

For a uniform nano-beam, the analytical solution for the governing equation of the torsional coordinate function writes as

$$\Theta(x) = \Upsilon_1 \sin\varsigma_1 x + \Upsilon_2 \cos\varsigma_1 x + \Upsilon_3 \sinh\varsigma_2 x + \Upsilon_4 \cosh\varsigma_2 x \tag{39}$$

where unknown integration constants $\Upsilon_k \ (k=1..4)$ have yet to be determined and $\varsigma_1, \varsigma_2$ can be expressed as

$$\varsigma_1^2 = \frac{(\rho\omega^2 L_c^2 - \mu) + \sqrt{(\mu - \rho\omega^2 L_c^2)^2 + 4\mu\rho\omega^2 \ell^2}}{2\mu\ell^2}$$

$$\varsigma_2^2 = \frac{-(\rho\omega^2 L_c^2 - \mu) + \sqrt{(\mu - \rho\omega^2 L_c^2)^2 + 4\mu\rho\omega^2 \ell^2}}{2\mu\ell^2} \quad (40)$$

For a fully-clamped nano-beam in the framework of modified nonlocal strain gradient theory, a homogeneous fourth-order algebraic system in terms of the unknown integration constants $\Upsilon_k \ (k=1..4)$ is established as a result of imposing kinematic BCs Eq. (26) along with CBCs Eq. (16) to the closed form solution of the torsional coordinate function Eq. (39). A similar homogeneous set of algebraic equations can be obtained for a fully-clamped nano-beam associated with nonlocal strain gradient theory by replacing the CBCs of NSG as Eq. (21). In the same way, homogeneous fourth-order algebraic systems can be determined for cantilever nano-beams consistent with either of nonlocal strain gradient models. It is well-known that to get a non-trivial solution, the resulted system of algebraic equations should be singular. Accordingly, a highly nonlinear characteristic equation will be obtained for fully-clamped and cantilever nano-beams associated with either of nonlocal strain gradient models that should be numerically solved.

Prior to presenting the numerical illustrations of torsional free vibrations, mode shapes and natural torsional frequencies of fully-clamped nano-beams corresponding to the nonlocal strain gradient theory are examined. It may be shown that the torsional coordinate function of NSG fully-clamped nano-beam coincides with the local mode shapes of torsional vibration for fully-clamped beams [31-32]. This controversial coincidence of torsional mode shapes will lead to the fundamental frequency of torsional free vibrations as [32]

$$\omega^2 = \left(\frac{\pi^2 J_\mu}{L^2 J_\rho}\right)\frac{1+\pi^2 \lambda_\ell^2}{1+\pi^2 \lambda_c^2} \qquad (41)$$

As it can be deduced form Eq. (41), the natural frequency of torsional free vibrations of NSG nano-beam coincides with the natural frequency of local beam for equal values of nonlocal and gradient parameters $\lambda_c = \lambda_\ell$. This controversial issue reveals another drawback of the classical nonlocal strain gradient model when applied to bounded continua of applicative interest. Unlike the nonlocal strain gradient theory, MNSG provides technically noteworthy size-dependent vibration responses.

5.1. *Numerical illustrations of torsional free vibrations*

Fundamental torsional frequencies of fully-clamped and cantilever nano-beams consistent with the modified nonlocal strain gradient theory are numerically detected and compared with those obtained by NSG. For consistency of illustrations, the non-dimensional fundamental torsional frequency $\bar{\omega}$ is defined by

$$\bar{\omega}^2 = \left(\frac{L^2 J_\rho}{\pi^2 J_\mu}\right)\omega^2 \qquad (42)$$

The illustrated fundamental torsional frequencies are also normalized utilizing their corresponding local natural frequencies $\bar{\omega}_{\text{LOC}}$. The variations of the normalized fundamental frequencies of fully-clamped and cantilever nano-beams consistent with MNSG and NSG in terms of the nonlocal and gradient characteristic parameters are demonstrated in Figs. 6 and 7, respectively. The small-scale characteristic parameters are assumed to have the same ranging set as the elastostatic torsional response illustrated in Figs. 2 and 3. It is noticeably deduced form the illustrative results that the nonlocal characteristic parameter $\lambda_c$ has the effect of decreasing the fundamental torsional frequencies, that is a larger $\lambda_c$ involves a smaller natural torsional frequency for a given value of $\lambda_\ell$. The natural torsional frequencies consistent with either of the nonlocal strain gradient models thus demonstrate a softening structural response in terms of the nonlocal characteristic parameter $\lambda_c$. Moreover, the natural frequencies associated with either of the nonlocal strain gradient models increase by increasing the gradient characteristic parameters $\lambda_\ell$, and accordingly, demonstrating a stiffening structural response in terms of the gradient characteristic parameter $\lambda_\ell$ for a given value of $\lambda_c$. The natural torsional frequencies of fully-clamped nano-beams associated with NSG exhibit a peculiar response to coincide with the fundamental frequencies of local beam as the small-scale characteristic parameters are identical. The torsional free vibration results in the framework of MNSG do not demonstrate such a controversial structural response of nonlocal strain gradient model. The fundamental torsional frequencies of the local elastic beam model can be also recovered as the nonlocal and gradient parameters approach zero $\lambda_c, \lambda_\ell \to 0^+$. The numerical values of normalized fundamental torsional frequencies of fully-clamped and cantilever nano-beams, detected in accordance with MNSG and NSG are reported in Tables 3 and 4, correspondingly.

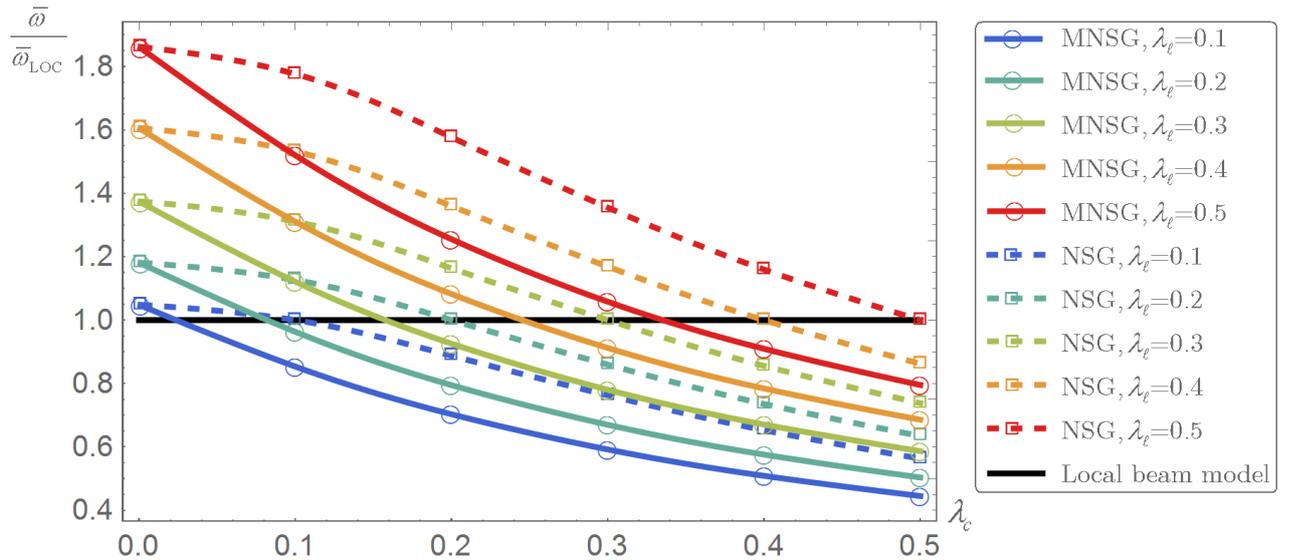

**Fig. 6.** Effects of characteristic parameters on the fundamental torsional frequency of fully-clamped nano-beams

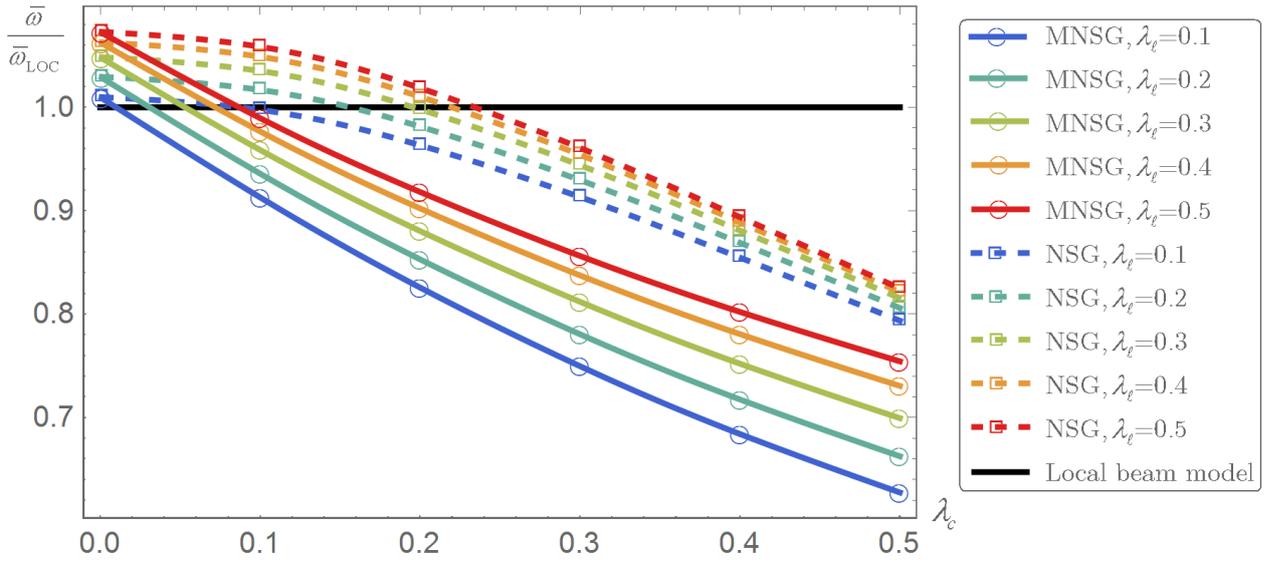

**Fig. 7.** Effects of characteristic parameters on the fundamental torsional frequency of cantilever nano-beams

# 6. Conclusions

Size-dependent torsional responses of elastic beams have been examined by making recourse to nonlocal strain gradient theory. The corresponding constitutive formulation of elastic torsion has been established by a variationally consistent approach, based on suitably selected test fields. Nonlocal convolution integrals governing the nonlocal strain gradient model have been replaced with a more convenient differential problem, equipped with new non-standard boundary conditions of constitutive type. Both elasto-static and –dynamic torsional responses of nano-beams have been investigated employing an efficient analytical solution procedure. Novel exact nonlocal solutions have been detected and compared with those obtained by the known model of nonlocal strain gradient elasticity available in literature that is equipped with inappropriate higher-order boundary conditions.

The main outcomes of the present study can be enumerated as follows.

- The need of prescribing the suitable form of (non-standard) constitutive boundary conditions for nano-beams has been highlighted.

- The contributions in literature on elastic torsion of nano-beams, in the framework of nonlocal strain gradient elasticity, have been amended to appropriately take the constitutive boundary conditions into account.

- The drawbacks due to imposing non-pertinent higher-order boundary conditions of strain gradient theory, as usually and widely adopted in literature, have been discussed.

- It has been demonstrated that the non-standard higher-order boundary conditions (usually adopted by the scientific community) have nothing to do with the appropriate variational formulation of nonlocal strain gradient elasticity developed in the present paper.

- The modified nonlocal strain gradient theory for elastic torsion, equipped with non-standard constitutive boundary conditions, leads to well-posed continuum problems.

- For elastic nano-beams subjected to uniformly distributed couples, it has been analytically shown that the known nonlocal strain gradient law is unable to capture nonlocal effects. On the contrary, the modified formulation of nonlocal strain gradient elasticity can efficiently exhibit both softening and stiffening behaviors which are peculiar in torsion of nano-beams of applicative interest.

- In both elasto-static and -dynamic analyses, a stiffening behavior has been revealed by MNSG for increasing gradient characteristic parameter and a softening response has been detected for increasing nonlocal characteristic parameter.

- The local elasticity model has been recovered as nonlocal and gradient parameters go to zero in the proposed formulation of modified nonlocal strain gradient elasticity.

- The results associated with torsional free vibrations of nano-beams also reveal another controversial issue of nonlocal strain gradient theory. Fundamental torsional frequencies of nonlocal strain gradient nano-beams coincide with those of corresponding local beams as nonlocal and gradient characteristic parameters are identical. This atypical structural behavior is absent in the modified nonlocal strain gradient model which amends thus previous formulations on the matter.

- Modified nonlocal strain gradient theory provides a simple and viable strategy for design and optimization of modern structural components of Nano-Electro-Mechanical-Systems.

**Acknowledgement**

Financial supports from the Italian Ministry of Education, University and and Research (MIUR) in the framework of the Project PRIN 2015 "COAN 5.50.16.01" - code 2015JW9NJT - and from the research program ReLUIS 2019 are gratefully acknowledged.

**Table 1.** Normalized maximum torsional rotation of fully-clamped nano-beam

$$\frac{\overline{\theta}_{max}}{\overline{\theta}_{max}^{LOC}}$$

| | MNSG | | | | | NSG | | | | |
|---|---|---|---|---|---|---|---|---|---|---|
| $\lambda_c$ | $\lambda_\ell = 0.1$ | $\lambda_\ell = 0.2$ | $\lambda_\ell = 0.3$ | $\lambda_\ell = 0.4$ | $\lambda_\ell = 0.5$ | $\lambda_\ell = 0.1$ | $\lambda_\ell = 0.2$ | $\lambda_\ell = 0.3$ | $\lambda_\ell = 0.4$ | $\lambda_\ell = 0.5$ |
| $0^+$ | 0.92108 | 0.73219 | 0.54261 | 0.39782 | 0.29611 | 0.921078 | 0.732183 | 0.542612 | 0.397814 | 0.296109 |
| 0.1 | 1.39461 | 1.13391 | 0.847537 | 0.623634 | 0.465042 | 0.921078 | 0.732183 | 0.542612 | 0.397814 | 0.296109 |
| 0.2 | 2.02599 | 1.66954 | 1.2541 | 0.924727 | 0.690288 | 0.921078 | 0.732183 | 0.542612 | 0.397814 | 0.296109 |
| 0.3 | 2.81521 | 2.33909 | 1.76231 | 1.30109 | 0.971844 | 0.921078 | 0.732183 | 0.542612 | 0.397814 | 0.296109 |
| 0.4 | 3.76227 | 3.14254 | 2.37216 | 1.75273 | 1.30971 | 0.921078 | 0.732183 | 0.542612 | 0.397814 | 0.296109 |
| 0.5 | 4.86718 | 4.0799 | 3.08365 | 2.27965 | 1.70389 | 0.921078 | 0.732183 | 0.542612 | 0.397814 | 0.296109 |

**Table 2.** Normalized mid-span torsional rotation of cantilever nano-beam

| | $\dfrac{\bar{\theta}_{x=1/2}}{\bar{\theta}^{LOC}_{x=1/2}}$ | | | | | | | | | |
|---|---|---|---|---|---|---|---|---|---|---|
| | MNSG | | | | | NSG | | | | |
| $\lambda_c$ | $\lambda_\ell = 0.1$ | $\lambda_\ell = 0.2$ | $\lambda_\ell = 0.3$ | $\lambda_\ell = 0.4$ | $\lambda_\ell = 0.5$ | $\lambda_\ell = 0.1$ | $\lambda_\ell = 0.2$ | $\lambda_\ell = 0.3$ | $\lambda_\ell = 0.4$ | $\lambda_\ell = 0.5$ |
| $0^+$ | 0.97370 | 0.91073 | 0.84754 | 0.79927 | 0.76537 | 0.973693 | 0.910728 | 0.847537 | 0.799271 | 0.76537 |
| 0.1 | 1.26487 | 1.17797 | 1.08251 | 1.00788 | 0.955014 | 0.973693 | 0.910728 | 0.847537 | 0.799271 | 0.76537 |
| 0.2 | 1.60866 | 1.48985 | 1.35137 | 1.24158 | 1.16343 | 0.973693 | 0.910728 | 0.847537 | 0.799271 | 0.76537 |
| 0.3 | 2.00507 | 1.84636 | 1.6541 | 1.50036 | 1.39061 | 0.973693 | 0.910728 | 0.847537 | 0.799271 | 0.76537 |
| 0.4 | 2.45409 | 2.24751 | 1.99072 | 1.78424 | 1.63657 | 0.973693 | 0.910728 | 0.847537 | 0.799271 | 0.76537 |
| 0.5 | 2.95573 | 2.6933 | 2.36122 | 2.09322 | 1.9013 | 0.973693 | 0.910728 | 0.847537 | 0.799271 | 0.76537 |

**Table 3.** Normalized fundamental torsional frequencies of fully-clamped nano-beam

| | $\dfrac{\bar{\omega}}{\bar{\omega}_{\text{LOC}}}$ | | | | | | | | | |
|---|---|---|---|---|---|---|---|---|---|---|
| | MNSG | | | | | NSG | | | | |
| $\lambda_c$ | $\lambda_\ell = 0.1$ | $\lambda_\ell = 0.2$ | $\lambda_\ell = 0.3$ | $\lambda_\ell = 0.4$ | $\lambda_\ell = 0.5$ | $\lambda_\ell = 0.1$ | $\lambda_\ell = 0.2$ | $\lambda_\ell = 0.3$ | $\lambda_\ell = 0.4$ | $\lambda_\ell = 0.5$ |
| $0^+$ | 1.04609 | 1.17865 | 1.37139 | 1.60276 | 1.85837 | 1.04818 | 1.181 | 1.37413 | 1.60596 | 1.86209 |
| 0.1 | 0.853416 | 0.963215 | 1.12131 | 1.31076 | 1.51996 | 1 | 1.12672 | 1.31097 | 1.53214 | 1.77649 |
| 0.2 | 0.702992 | 0.794818 | 0.92573 | 1.08234 | 1.2552 | 0.887535 | 1 | 1.16353 | 1.35983 | 1.5767 |
| 0.3 | 0.591869 | 0.669714 | 0.78017 | 0.912219 | 1.05795 | 0.762794 | 0.859453 | 1 | 1.16871 | 1.3551 |
| 0.4 | 0.508701 | 0.575719 | 0.670681 | 0.784199 | 0.909472 | 0.652682 | 0.735388 | 0.855646 | 1 | 1.15948 |
| 0.5 | 0.444903 | 0.50346 | 0.586464 | 0.685706 | 0.795233 | 0.562907 | 0.634237 | 0.737954 | 0.862452 | 1 |

**Table 4.** Normalized fundamental torsional frequencies of cantilever nano-beam

| | $\dfrac{\bar{\omega}}{\bar{\omega}_{\text{LOC}}}$ | | | | | | | | | |
|---|---|---|---|---|---|---|---|---|---|---|
| | MNSG | | | | | NSG | | | | |
| $\lambda_c$ | $\lambda_\ell = 0.1$ | $\lambda_\ell = 0.2$ | $\lambda_\ell = 0.3$ | $\lambda_\ell = 0.4$ | $\lambda_\ell = 0.5$ | $\lambda_\ell = 0.1$ | $\lambda_\ell = 0.2$ | $\lambda_\ell = 0.3$ | $\lambda_\ell = 0.4$ | $\lambda_\ell = 0.5$ |
| $0^+$ | 1.00881 | 1.02852 | 1.04746 | 1.06188 | 1.07204 | 1.00981 | 1.02951 | 1.04841 | 1.0628 | 1.07294 |
| 0.1 | 0.912866 | 0.935842 | 0.95874 | 0.976675 | 0.98954 | 0.997551 | 1.01682 | 1.03513 | 1.04893 | 1.05858 |
| 0.2 | 0.825677 | 0.852951 | 0.880371 | 0.902234 | 0.91814 | 0.963275 | 0.981348 | 0.998029 | 1.01026 | 1.01862 |
| 0.3 | 0.749629 | 0.780515 | 0.811928 | 0.837437 | 0.856273 | 0.91323 | 0.929571 | 0.943966 | 0.954072 | 0.960761 |
| 0.4 | 0.68401 | 0.717446 | 0.752015 | 0.780641 | 0.802117 | 0.854659 | 0.869001 | 0.880877 | 0.888776 | 0.893806 |
| 0.5 | 0.627475 | 0.662483 | 0.69937 | 0.730533 | 0.754293 | 0.793623 | 0.805925 | 0.815388 | 0.821315 | 0.824941 |